\documentclass[10pt,a4paper]{article}
\usepackage{epsfig}
\usepackage{graphicx}               
\usepackage{amsmath}                
\usepackage{lineno}
\usepackage{url}

\title{Hybrid ECAL: Optimization and Related Developments}

\author{T.~Suehara\thanks{
Speaker. Talk presented at the International Workshop on Future Linear Colliders (LCWS14), Belgrade, Serbia, 6-10 October 2014.}, ~H.~Hirai, H.~Sumida, H.~Ueno, \\
Y.~Sudo, T.~Yoshioka and K.~Kawagoe \\[+2mm]
\small{Kyushu University, 6-10-1 Hakozaki, Higashi-ku, Fukuoka, 812-8581 Japan} \\[+2mm]
}
\date{\today}

\begin{document}
\maketitle

\begin{abstract}

`Hybrid ECAL' is a cost-conscious option of electromagnetic calorimeter (ECAL) for particle flow calorimetry
to be used in a detector of International Linear Collider (ILC).
It is a combination of silicon-tungsten ECAL, which realizes high granularity and robust measurement of electromagnetic
shower, and scintillator-tungsten ECAL, which gives affordable cost with similar performance to silicon.
Optimization and a data acquisition trial in a test bench for the hybrid ECAL are described in this article.

\end{abstract}

\section{Introduction}

International Linear Collider (ILC) is a next-generation electron-positron linear collider.
International Large Detector (ILD) is one of two validated detector concepts for the ILC.
ILD is based on particle flow concept, which requires separation of each particle in jets.
Jet energy in the particle flow is measured with momentum of tracks and energy of neutral clusters.
Since momentum resolution of tracks is much better than energy resolution of clusters in general,
this gives better energy resolution of jets.

The particle flow requires a high granular calorimeter system to separate particles, especially in
electromagnetic calorimeter (ECAL). ILD ECAL has $5\times5$ mm$^2$ granularity, which is a major challenge
in both detector and electronics, and requires high cost. Two options are available for ILD ECAL:
one is silicon-tungsten ECAL (SiECAL) and the other is scintillator-tungsten ECAL (ScECAL).
SiECAL utilizes silicon diode pads for readout, sandwiched with tungsten absorber.
Silicon pads can be easily divided to desired area ($5\times5$ mm$^2$)
and the total cost mainly depends on the sensor area.
ScECAL utilizes strip scintillators with silicon-photomultipliers (SiPMs) for readout instead of silicon.
To reduce number of SiPM sensors which are the cost driver,
strip scintillators ($5\times45$ mm$^2$) placed perpendicularly to the neighbor layers have been adopted for the ScECAL design.
The total cost of ScECAL is around a half of SiECAL in the cost estimation in Detailed Baseline Design (DBD)~\cite{Behnke:2013lya} report.
The average cost of SiECAL and ScECAL is around 30\% of total cost of ILD
with the baseline design of inner radius of ECAL at 1800 mm from the center and 30 layers of sensors.

\begin{table}[htb]
  \begin{center}
    \begin{tabular}{|l|l|l|}\hline
        & SiECAL & ScECAL \\ \hline\hline
 sensor & silicon & scintillator with SiPM \\ \hline
 pixel size & $5\times5$ mm$^2$ & $5\times45$ mm$^2$ \\ \hline
 thickness & 320 to 500 $\mu$m & 1 or 2 mm \\ \hline
 MIP response & around 25 K pairs       & $\mathcal{O}(10)$ photoelectrons \\
              & in 320 $\mu$m thickness & (depending on detailed structure) \\ \hline
 sensor gain & 1 & $\mathcal{O}(10^5)$ \\ \hline
 sensor stability & stable & varied by temperature and \\
                 &         & sensor overvoltage \\
                 &         & periodic calibration required \\ \hline
 gain in electronics & higher & lower \\ \hline
 saturation & only in electronics & SiPM saturation, depending on \\
            &                    & number of pixels of SiPM\\ \hline
 assembly & easier & complicate, including wrapping \\ 
          &        & and placing tiny strips \\ \hline
 cost & higher & lower \\ \hline
    \end{tabular}
    \caption{Comparison of SiECAL and ScECAL.}
    \label{tbl:compsisc}
  \end{center}
\end{table}

Table \ref{tbl:compsisc} shows comparison of characteristics of SiECAL and ScECAL.
Each has advantages and disadvantages. SiECAL has advantages on number of pairs,
sensor stability and granularity, though requirements on electronics are higher
(number of channels and gain of preamplifier) and sensor cost is higher.
ScECAL is a cheaper solution but has stronger requirements
on periodic sensor calibration, saturation correction and assembly.

To keep the DBD cost equal or lower with using SiECAL, we have two options:
(1) shrinking detector to smaller size, with reduced number of sensors,
(2) introducing hybrid ECAL, which is a combination of SiECAL and ScECAL.

The option (1) keeps robustness and simplicity of SiECAL but the detector
performance should be degraded at some extent because of worse particle
separation and worse momentum resolution of tracks due to the smaller detector.
The performance degradation should be carefully estimated since the degradation
should usually be compensated by more luminosity to obtain the similar impact of
physics results. More luminosity means more operation cost, which may be equal or
more to the reduction of detector cost.

With the option (2) performance degradation should be much smaller than option (1)
but more complexity is introduced. One consideration is that one of the HCAL option
of ILD is also scintillator-SiPM complex. If it is adopted, the total complexity of
calorimeter system is maintained at similar level compared to the simple SiECAL and
scintillator HCAL. In the hybrid ECAL, the SiECAL technology should be used in the inner 
part of ECAL, which requires more granularity to separate particles and then expect
more ghost hits if we use strip ScECAL sensors.

\section{Optimization of hybrid ECAL}

Studies are ongoing to establish a optimal hybrid ECAL configuration in ILD.
We consider optimization with constraints that: (1) ILD of similar size to DBD version,
(2) ILD of similar cost to DBD and (3) combination of SiECAL and ScECAL.
Parameters of the optimization include (a) number of layers
and thicknesses of absorber, (b) pixel size of each layer, and
(c) order and fraction of SiECAL and ScECAL.

\begin{figure}[htbp]
  \begin{center}
    \includegraphics[width=0.7\linewidth]{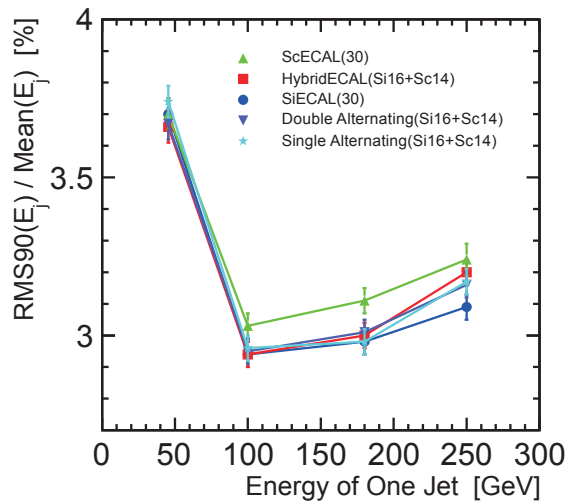}
  \end{center}
  \caption{Comparison of jet energy resolution among non-alternating (`HybridECAL', inner SiECAL and outer ScECAL),
single and double alternating configuration of hybrid ECAL, full SiECAL and ScECAL configuration.}
  \label{fig:alternate}
\end{figure}

For (c), resolution of jet energy at various energies has been compared
with SiECAL, ScECAL and several order of hybrid ECAL in Fig.~\ref{fig:alternate}.
ILD full simulation and reconstruction software (ILCSoft~\cite{ILCsoftweb} v01-16-02 with Pandora PFA~\cite{Thomson:2009rp} v00-09-02)
with $q\bar{q}$ two jet events are used.
The three configurations of normal, single alternating and double alternating
give almost same jet energy resolution at every energy.

For (a), a comprehensive study is planned to understand response to longitudinal parameters
of sandwiched calorimeter. Longitudinal configuration of ILD ECAL is separated to two: inner and outer layer.
Current baseline adopts the configuration with thickness ratio of 1 by 2 in inner and outer layer,
separated by the half of radiation length with total 29 (inner 20 and outer 9) layers with additional
one layer before absorber. Total absorber thickness is set to 22.8 radiation length.

However, the thickness ratio, total number of layers and the border position of inner and outer layer
is practically not optimized. We can also consider three configuration of inner, middle, outer or 
even fully variant thickness in each layer.

In the jet energy measurements with the particle flow algorithm, the longitudinal structure mainly affects
the intrinsic energy resolution of single particle, while the transverse structure is more important on the
clustering to separate contribution of charged particles.
Contribution of resolution of single particle in ECAL is not large in jet energy measurements since
HCAL energy resolution dominates the performance on the lower energy and the clustering dominates
on the higher energy.
However, ILC features non-jet measurements as well as jet physics, such as $\pi_0$ reconstruction from
$\tau$ decay, $H \to \gamma\gamma$ and non-pointing photons from new physics models.
These measurements heavily depend on the ECAL energy resolution and thus longitudinal structure is still important.

\begin{figure}[htbp]
  \begin{center}
    \includegraphics[width=0.7\linewidth]{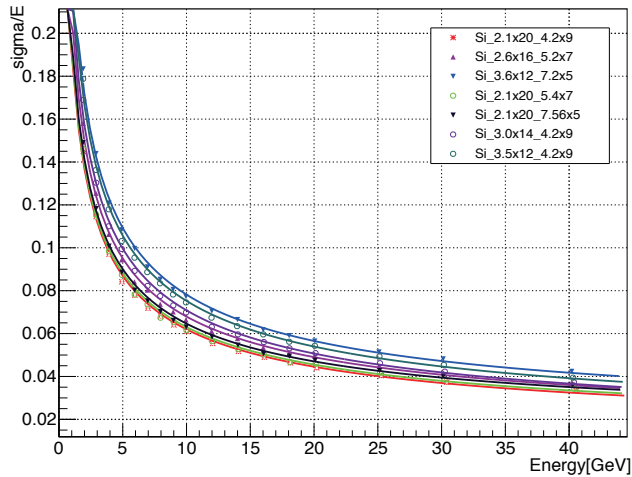}
  \end{center}
  \caption{Comparison of energy resolution of single photons, by reducing number of layers in inner and outer region.}
  \label{fig:photon}
\end{figure}

As the first step of optimization, we compared energy resolution of single photons with several longitudinal configuration,
shown in in Fig.~\ref{fig:photon}. In this figure, degradation of photon energy resolution is larger with reducing inner
layers than reducing outer layers.

A lot of analysis are still needed to optimize the hybrid ECAL, such as looking at hadron energy resolution,
confusion study with various transverse configuration, etc.
We plan to conclude one hybrid model to be compared with small detector after the optimization studies.

\section{Combined DAQ}

Hardware aspect is as important as the optimization aspects.
CALICE~\cite{CALICEweb} collaboration aims to combine all efforts of linear collider calorimeter development into a consistent manner.
For the calorimeter data acquisition (DAQ), many detector systems for ILC, including silicon and scintillator system, are based on
`ROC'-family integrated readout chip developed at OMEGA~\cite{OMEGAweb} group. The all ROC chips are based on a same basic design,
and have similar protocol for configuration and data readout.

Interface of ROC chips, timing control and data readout has been also developed for the all `ROC'-based
CALICE systems, but each detector system currently uses readout structure
which is not fully compatible each other by historical reasons.
To integrate full calorimeter system (and more), we need to develop a consistent
readout system. In hybrid ECAL system, we are trying to develop a combined DAQ system of
SiECAL and ScECAL, which can be a good start point to larger integration.

Our design is to use existing hardware and software in each system, with minimal interaction of them.
The interaction includes clock synchronization, common readout counting, run and stop controller software
working at higher level than each readout system, and combined data collection software.

\begin{figure}[htbp]
\begin{center}
\begin{minipage}{0.48\linewidth}
\includegraphics[width=1\linewidth]{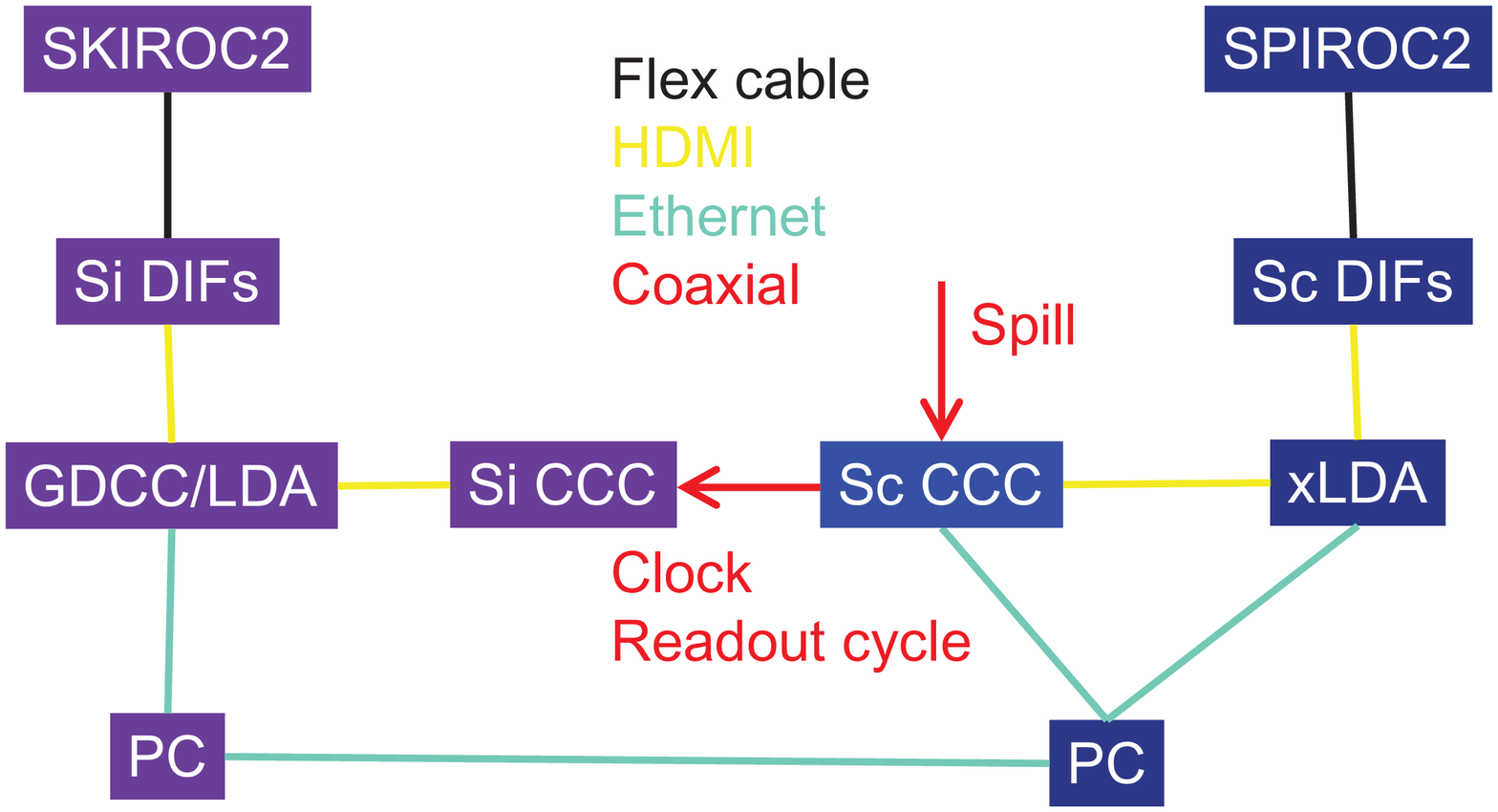}
\end{minipage}
\begin{minipage}{0.48\linewidth}
\includegraphics[width=1\linewidth]{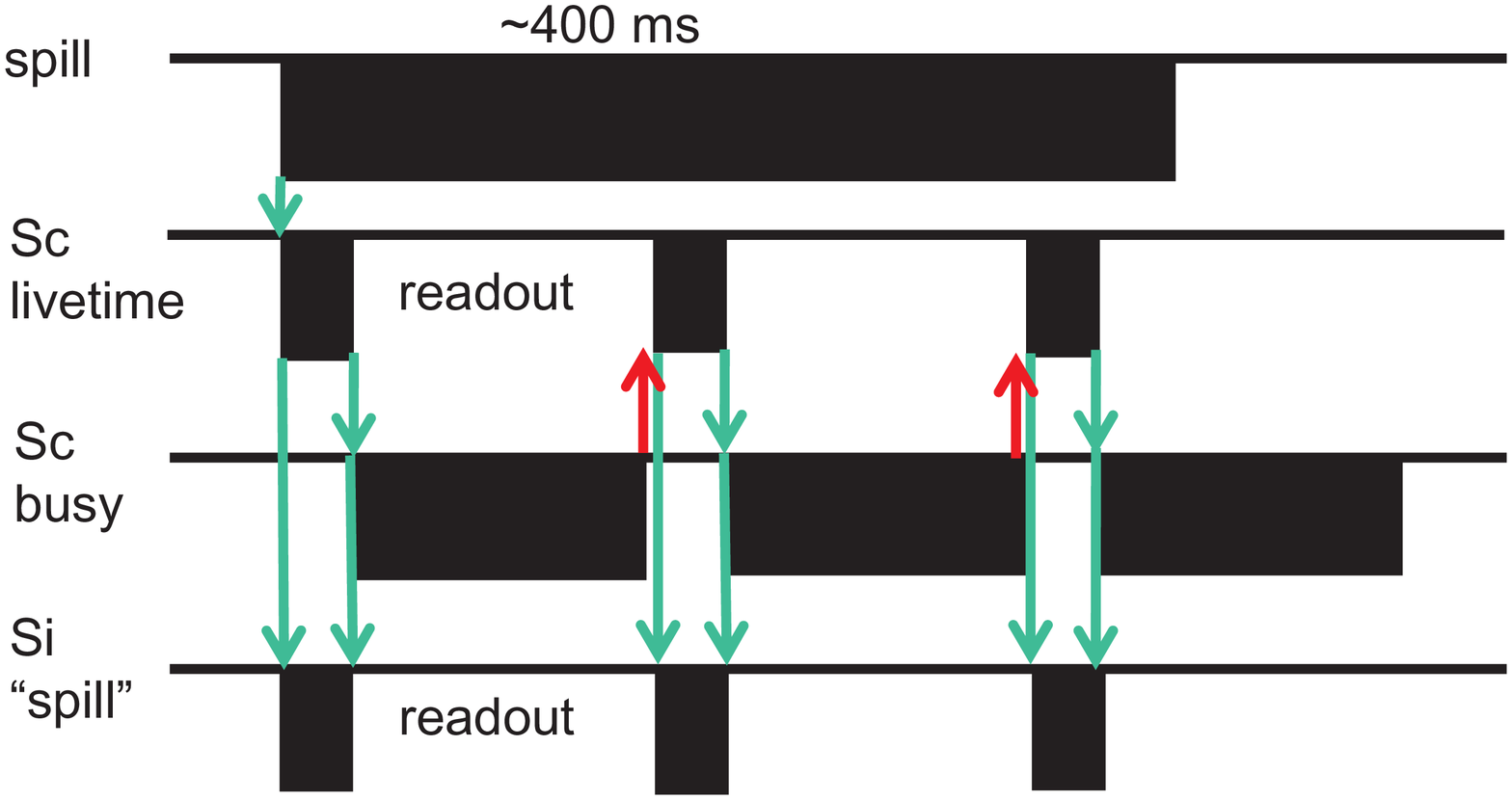}
\end{minipage}
\end{center}
\caption{Block diagram and timing chart of combined DAQ.}
\label{fig:hardware}
\end{figure}

Figure \ref{fig:hardware} shows the block diagram and timing chart of the combined DAQ hardware.
`Spill' in the figures comes from testbeam control. The spill in the testbeam is assumed to be
different from ILC operation mode, that long spills (400 ms in the timing chart) come
several times a minute (ILC spill is only 1 ms long with 5 Hz operation).
To maximize data taking efficiency, the spill should be subdivided to several `readout cycle's, determined by
scintillator CCC (Clock and Control Card) by looking at `busy' signals from each scintillator module.
Scintillator busy is flagged from each SPIROC2 chip when the memory of the chip is full and
cleared when the readout has finished (silicon busy is not treated due to a technical reason).
The acquisition is stopped with busy flag, and the next readout cycle is started
after all busy flags are cleared. The acquisition period is shared from scintillator CCC
to silicon CCC by a level signal, provide the same readout cycle counting.
Scintillator CCC also creates a master clock to be synchronized with silicon CCC.

\begin{figure}[htbp]
  \begin{center}
    \includegraphics[width=0.48\linewidth]{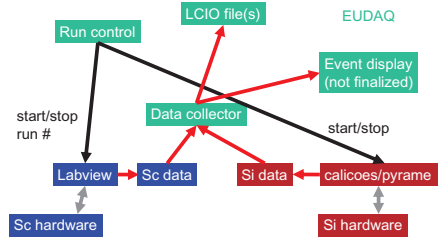}
  \end{center}
  \caption{Software structure of combined DAQ.}
  \label{fig:software}
\end{figure}

Figure \ref{fig:software} shows the software structure.
SiECAL and ScECAL have their own readout software, based on a system called Calicoes~\cite{Cornat:2014cha} in SiECAL
and a LabVIEW-based system in ScECAL. For the combined system, we adopted EUDAQ~\cite{EUDAQweb} framework,
which can be connected to each readout system via TCP socket connection.
We developed a EUDAQ producer, which talks to each readout system via socket to obtain raw data packets.
The received raw data is converted to LCIO format within the EUDAQ framework, to be stored in LCIO files.
Readout cycle is numbered and checked in the EUDAQ data collector to ensure that
the same event data of SiECAL and ScECAL is stored in the same LCIO event.
The EUDAQ run control is also used to provide start and stop signal to each system and assign run numbers.

The combined DAQ has been tested at CERN PS testbeam facility from November 26 to December 8, 2014.
A SiECAL layer was placed in front of a scintillator stuck, including three ScECAL layers and eleven HCAL layers.
The combined data taking run successfully, taking data with 7 GeV muons and 2-8 GeV pions.
The concurrent hits have been found between silicon and scintillator layers,
with consistent timing difference of electronics, proving successful synchronization of two systems.
Detailed analysis is ongoing.

\section{Summary}

Hybrid ECAL is an cost-effective option for ILD ECAL. We have started the optimization of hybrid ECAL
by looking at energy resolution of photons, hadrons and jets with various configuration.
The alternation of SiECAL and ScECAL layers gives similar performance to the combination of
SiECAL in inner part and ScECAL in outer part. Reducing number of layers have some impact on the
photon energy resolution, and the inner region is more important.
For the DAQ, we developed a combined SiECAL and ScECAL DAQ with minimal modification
of each readout framework.
We adopted EUDAQ as a higher level run control and data integration.
The testbeam at CERN was successful, obtaining concurrent hits at silicon and scintillator layers.

\section*{Acknowledgment}

We thank LLR, DESY, Shinshu and CERN colleagues for supplying sensors, DAQ and testbeam environment.
The computing resource for the optimization study was mainly provided from KEK computing center.
This work was supported by MEXT/JSPS KAKENHI Grant Numbers 23000002 and 23104007,
and by Kyushu University Interdisciplinary Programs in Education and Projects in Research Development.

\bibliographystyle{plain}
\bibliography{ref}

\end{document}